\documentclass{bmcart}

\usepackage{amsthm,amsmath}
\usepackage[utf8]{inputenc} 

\usepackage{graphicx}
\usepackage{siunitx}

\startlocaldefs
\endlocaldefs

\begin{document}

\begin{frontmatter}

\begin{fmbox}
\dochead{Research}


\title{Burstiness of human physical activities and their characterisation}


\author[
   addressref={aff1, aff3},                   
   corref={aff1},                       
   noteref={n1},                        
   email={takemak@gmail.com}   
]{\inits{MT}\fnm{Makoto} \snm{Takeuchi}}
\author[
   addressref={aff2}
]{\inits{YS}\fnm{Yukie} \snm{Sano}}


\address[id=aff1]{
  \orgname{Policy and Planning Sciences, University of Tsukuba}, 
  \city{Ibaraki},                              
  \cny{Japan}                                    
}
\address[id=aff2]{%
  \orgname{Faculty of Engineering, Information and Systems, University of Tsukuba},
  \city{Ibaraki},
  \cny{Japan}
}
\address[id=aff3]{%
  \orgname{CyberAgent, Inc.},
  \street{Abema Towers 40-1 Udagawacho Shibuya-ku},
  \postcode{150-0042}
  \city{Tokyo},
  \cny{Japan}
}


\begin{artnotes}
\note[id=n1]{Equal contributor} 
\end{artnotes}

\end{fmbox}


\begin{abstractbox}

\begin{abstract} 
Human behaviour is heterogeneous and temporally fluctuates. 
Many studies have focused on inter-event time (IET) fluctuations and have reported that the IET distributions have a long-tailed distribution, which cannot be explained by a stationary Poisson point process. Such phenomenon observed in IET distributions are known as burstiness. Burstiness has also been reported for human physical activity, but the mechanism underlying it has not been clarified.
In this study, we collected human physical activity data while specifying the age of the subjects and their situations (for example, children's play and adults' housework), and we analysed their event time-series data.
We confirmed the burstiness in both children and adults. For the first time, burstiness studied in physical activities of children between the ages 2 and 5. We also confirmed that the characteristics of the IET distribution are unique to each activity situation. 
Our results may be critical in the identification of the burstiness mechanisms in human physical activity.

\end{abstract}




\end{abstractbox}
%

\end{frontmatter}



\section{Introduction}
\label{sec:1}
Temporal ﬂuctuations in dynamic patterns are studied as a typical phenomenon in many complex systems~\cite{karsai2018bursty}. Particularly, it has been reported that the distribution of inter-event times (IETs) in a system frequently displays a long-tailed distribution rather than an exponential distribution, such as the power-law distribution. This feature is referred to as burstiness. Burstiness suggests that these behaviours are not randomly generated by a stationary Poisson process; conversely, they are generated by a nontrivial mechanism that causes long eventless time periods and short periods of consecutive event occurrences~\cite{Barabasi2005}.

Burstiness has been observed in earthquakes~\cite{PhysRevLett.92.108501} and solar flares~\cite{ROSS2020124775} as a natural phenomenon. Burstiness has also been reported in neuronal activity~\cite{10.1162/089976603322518759}, animal activity~\cite{REYNOLDS2011245}, and urban soundscape~\cite{DESOUSA2019121557}. In addition, burstiness has been observed in various human behaviour contexts and hierarchies, such as e-mailing~\cite{Barabasi2005,PhysRevE.73.036127}, making phone calls~\cite{PhysRevE.83.025102}, web chatting~\cite{ZHANG2020122854}, blogging~\cite{YAN2017775}, website browsing~\cite{PhysRevE.73.036127}, book loaning~\cite{PhysRevE.73.036127,LEE2021125473},  stock trading~\cite{PhysRevE.73.036127}, and word occurrences in texts~\cite{CUI2017103}.
Furthermore, human physical movements, such as daily activities for several days~\cite{PhysRevLett.99.138103}, walking~\cite{PICOLI2022127160}, and touching a smartphone screen~\cite{PhysRevE.102.012307} also exhibit burstiness.

Whether the burstiness observed in these various complex systems is the result of a common mechanism is an intriguing question. 
Particularly in human behaviour, it is known that the burstiness of a particular type of behaviour has universality, i.e., the power-law exponents of an individual's power-law distribution have the same value~\cite{PhysRevE.73.036127}. 
Meanwhile, it has been reported that, depending on the presence or absence of a depressive illness, there exists differences in the power-law exponents of the distribution of IETs for physical activities ~\cite{PhysRevLett.99.138103}. Additionally, there are individual differences in the power-law exponents of the distribution of IETs for smartphone touch events~\cite{PhysRevE.102.012307}.
This apparent discrepancy could be the result of the differences in the population and behaviour types being analysed. 
Human physical activity involves different kinds of behaviors and requires a different interpretation of what is happening in the IETs period depending on the length of the time scale of focus. These are features that differ from a single type of behavior, such as sending an email. Due to this background, it is non-trivial and not well-tested whether the existing mechanisms that explain burstiness can also be applied to human physical activities.

This study provides insight into the origins of burstiness in human behaviour, particularly physical activity, through an experimental approach. In particular, we focus on what causes differences in burstiness. We collected human physical-activity data while clarifying the attributes of the subjects (children or adults) and their activity situations. We then identified burstiness and a unique long-tailed curve in the distribution of IETs for each situation. The finding indicates that the burstiness of human physical activity is not caused by a Poisson process that changes the event rates over time. These ﬁndings contribute to the research on the mechanism of human  burstiness. 

\section{Methodology}
First, acceleration time-series data on physical activity were obtained using an accelerometer attached to the subject's body. Second, we calculated the activity time series from the acceleration time series data and converted it into the event time series. Finally, we computed statistical features, such as burstiness, using the obtained event time series.

\subsection{The sensor and data collection}
We used the \textit{Mono Wireless TWELITE 2525A} accelerometer sensors with a measurement range of $\pm 157~\si{\metre\square\per\second}$. 
The antenna on this sensor transmits measurement data to a receiver, such as a computer or smartphone. It measures acceleration in three-axial directions at a sampling rate of $10~\si{\hertz}$. This sensor weighs $6.5~\rm{g}$, including a coin cell battery, and is small ($25~\si{\milli\metre}\times25~\si{\milli\metre}\times10~\si{\milli\metre}$). 
The sensor was placed in a custom-made case inaccessible to a child, and was attached to their waist using a belt.

Twelve subjects were measured, including three children aged between 2 and 3, and nine adults aged between 20 and 40. We obtained consent from the subjects before collecting the data. We obtained parental consent to collect data from their children. Our measurements for children and adults were performed indoors for 90 minutes (min.) to avoid wearing a sensor for long duration, particularly for children. Parents were present when their children's measurements were taken. The conditions during the measurement, such as during housework, desk work, and rest, were also recorded. The data were measured  five times for each situation. This study was approved by the Ethics Review Committee of the Faculty of Engineering, Information and Systems at the University of Tsukuba (\# 2019R295).

\subsection{Data pre-processing}
Missing data or ﬂuctuations in the measurement interval are possible owing to the accelerometer's data transmission process. Therefore, we excluded the data in which one or both of the following conditions occurred during the 90-min. observation period.
\begin{itemize}
    \item The interruption lasted longer than a minute.
    \item There are more than 30 interruptions that lasted 10 seconds or longer.
\end{itemize}

\begin{figure}[ht]
\centering
 \includegraphics[scale=0.3]{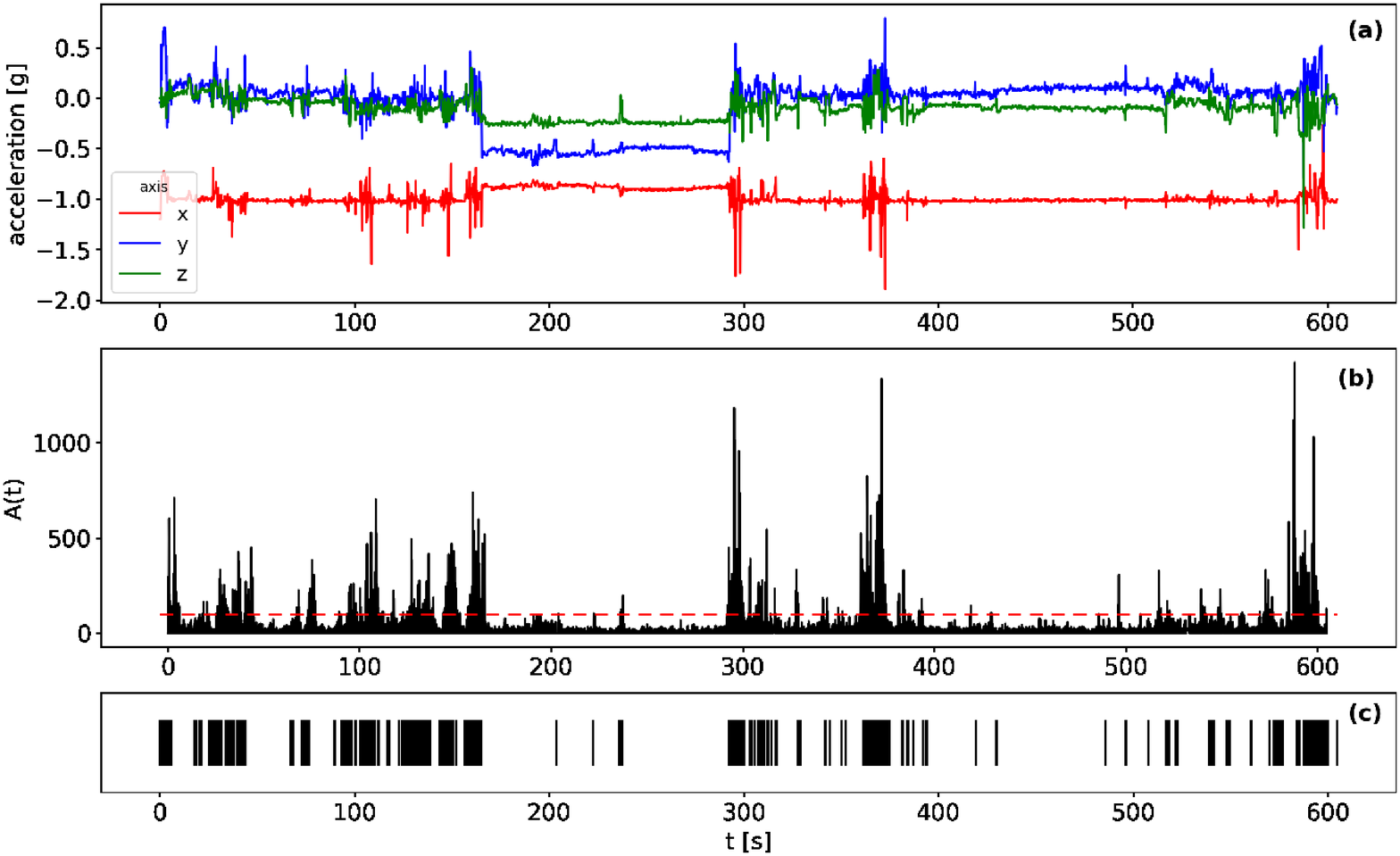}
 \caption{Event series data obtained from the pre-processing. (a) A 10 min. sample of three-dimensional acceleration time-series data on adult housework situations. (b) Time series data of physical activity, $A(t)$, and a threshold, $A_c$, for determining activity events. (c) Event series data, $E(t)$, created as activity events when the amount of activity exceeds a threshold value.}
 \label{fig:preprocessing}
\end{figure}

After data cleaning, we obtained the three-axes acceleration data (Fig.~\ref{fig:preprocessing}(a)).
Then the amount of activity $A(t)$ was calculated from the three-axis acceleration data as follows:

\begin{eqnarray}
A(t) = (\sqrt{\Delta x(t)^2+\Delta y(t)^2+\Delta z(t)^2})/\Delta t
\label{eq:a}
\end{eqnarray}
where $\Delta x(t), \Delta y(t),$ and $\Delta z(t)$ are the amount of change in acceleration in the three axes at $t$, respectively.
The observation's time interval, $\Delta t$, is approximately 100 milliseconds~$(\si{\milli\second})$; however, $\Delta t$ may vary for each data point owing to measurement-related ﬂuctuations. 

To compensate for the effects of this $\Delta t$ fluctuation, $A(t)$ is calculated by dividing the change in acceleration on the three-axes by $\Delta t$.
Figure~\ref{fig:preprocessing}(b) shows the time series of $A(t)$ calculated using the three-axes acceleration data as shown in  Fig.~\ref{fig:preprocessing}(a).

Next, the activity time series data $A(t)$ are converted to event-series data $E(t)$. An activity event is considered to have occurred when $A(t)$ exceeds a specific threshold value $A_c$.
\begin{equation}
    E(t) = \left\{
    \begin{array}{ll}
1 & (A(t) \geq A_c)\\
0 & (A(t) < A_c)
\end{array}
\right.
\end{equation}
Here, we adopted a constant value for this threshold, which is independent of individuals. 
We empirically set the threshold value $A_c=100~ \si{\metre\cubic\per\second}$. 
Additionally, we confirmed that the IET distribution results obtained in this study are typically independent of the threshold $A_c$ within certain ranges (see APPENDIX).

Finally, we computed the sequence of IETs from $E(t)$. 
There is a minimum limit to the time resolution of the sensor.
For example, if we observe an active state for more than $100~\si{\milli\second}$ at a sampling rate of $10~\si{\hertz}$, an event is observed every $100~\si{\milli\second}$, resulting in multiple consecutive $100~\si{\milli\second}$ IETs. 
However, these are pseudo-IETs produced by the sensor's time resolution; events do not actually continue to occur at $100~\si{\milli\second}$ intervals. 
For this purpose, we excluded these pseudo-IETs in the computation of the statistics. Because the sensor data are not always obtained at exactly $10~\si{\hertz}$ and contains fluctuations, this study excluded IETs of less than $300~\si{\milli\second}$.
In other words, the IETs of less than $300~\si{\milli\second}$ were assumed to be unobservable although they occurred, and they were treated as event-occurrence states rather than intervals. The IETs obtained from each 90 min. observation were combined for five observations per situation, and their distribution was investigated.

\subsection{Data analysis}
Statistical features of burstiness and temporal correlation were computed for the event time-series data, $E(t)$.
For burstiness, we evaluated the distribution of IETs and the burstiness parameter~\cite{Goh_2008, PhysRevE.94.032311}.
Here we used burstiness parameter $B_n$ with correction, by the number of events $n$ as follows~\cite{PhysRevE.94.032311}:

\begin{eqnarray}
B_n = \frac{\sqrt{n+1} \left( \frac{\sigma}{\langle \tau \rangle} \right)-\sqrt{n-1}}{(\sqrt{n+1}-2)\left(\frac{\sigma}{\langle \tau \rangle}\right)+\sqrt{n-1}}
\label{eq:bn}
\end{eqnarray}
where $n, \sigma$, and $\langle \tau \rangle$ denote the number of events, the standard deviation, and the mean of IETs, respectively.

\section{Results}
\subsection{IET distribution}
\begin{figure}[ht]
\centering
 \includegraphics[scale=0.5]{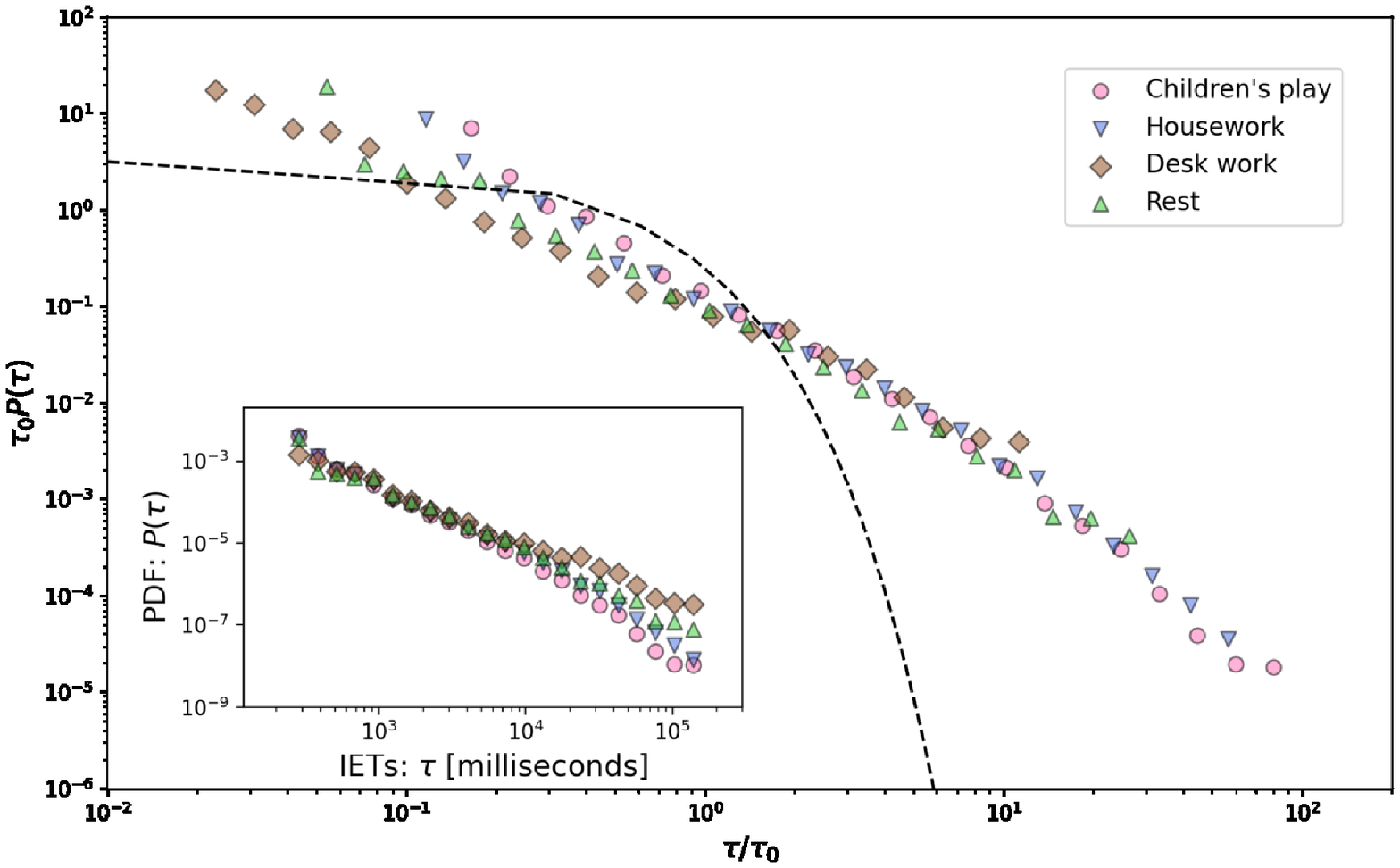}
 \caption{$\tau_0$ scaled IET distributions are shown for each activity situation type in log-log plot. Unscaled IET distributions are shown in the inset. The exponential distribution is shown as a dashed line for reference. Each situation is plotted using different marks.}
 \label{fig:iets_all}
\end{figure}

Figure~\ref{fig:iets_all} shows the distribution of IETs by situation, including \textit{Children's play}. The IET distributions are plotted by combining the five measurements for each situation. Here, we show the distributions scaled using the mean $\tau_0$ of individual IETs~\cite{Goh_2008}. All distributions show long-tail rather than exponential decay and do not overlap even for the $\tau_0$ scaled curves.
This study is the ﬁrst to confirm the burstiness in children's activities.

According to \cite{10.1371/journal.pone.0085777, doi:10.1137/070710111}, the goodness of ﬁt of several candidate distribution functions was evaluated for each situation to identify the distribution function in each situation. The candidate distribution functions include the exponential, stretched exponential, lognormal, power-law, and truncated power-law. Here, the truncated power-law behaves as power-law scaling in a certain range and is truncated using an exponentially bounded tail and expressed as shown in Eq.~(\ref{eq:truncated_power_law}).

\begin{eqnarray}
P(x) \propto x^{-\alpha} \exp{\left(-\lambda x\right)}
\label{eq:truncated_power_law}
\end{eqnarray}

\begin{figure}[ht]
\centering
 \includegraphics[scale=0.5]{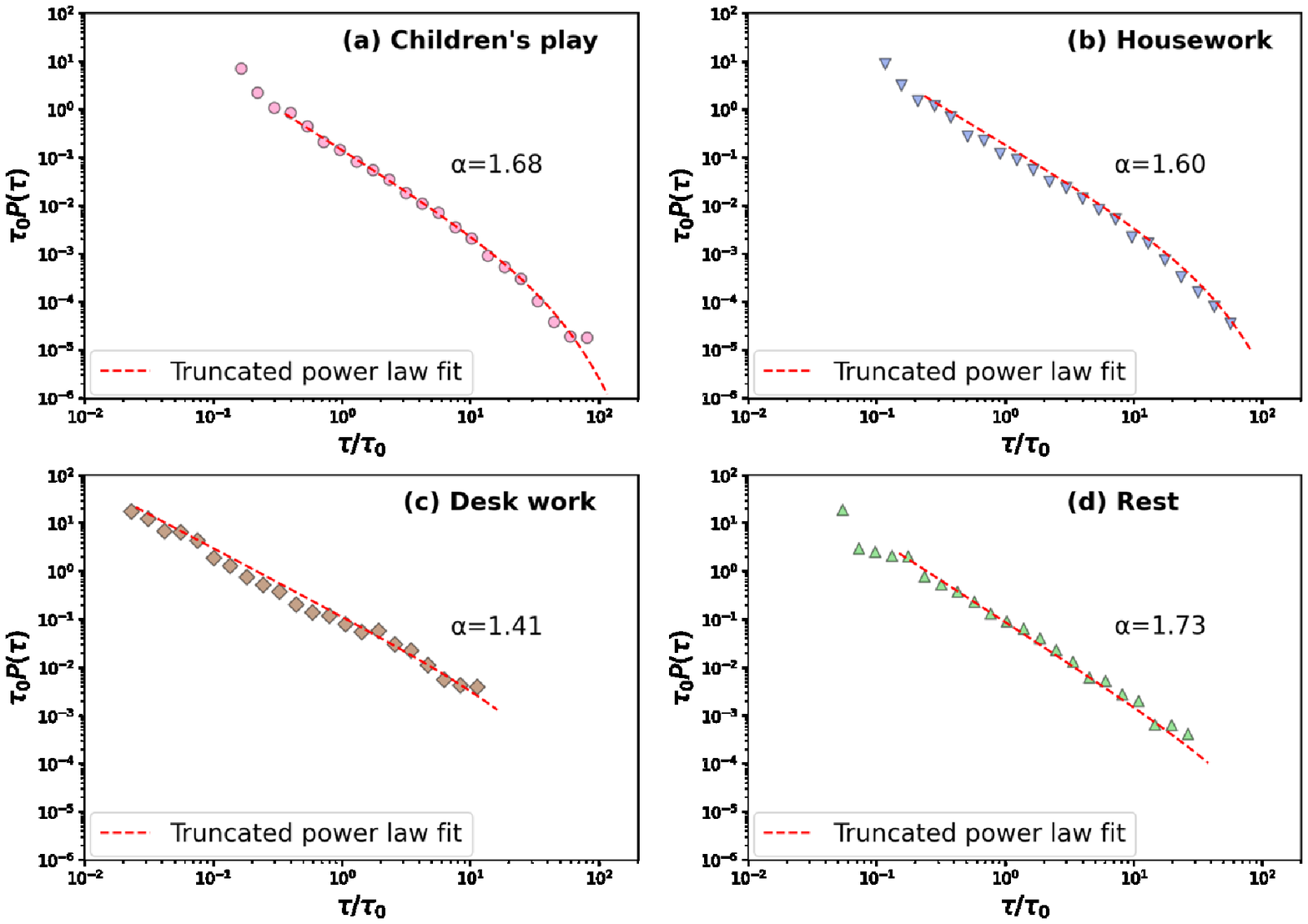}
 \caption{The results of fitting the truncated power-law by maximum likelihood estimation for each situation.}
 \label{fig:iets_dist_by_type}
\end{figure}

The evaluation results demonstrated that the truncated power-law distribution was the best function for all distributions. Figure~\ref{fig:iets_dist_by_type} and Table~\ref{table:result} show the results of ﬁtting the observed data for each situation to the truncated power-law using maximum likelihood estimation.

To determine whether the obtained curves are specific to each situation, we further examined the differences in the distributions using the Mann–Whitney U test with two-sided alternative hypothesis. The results show that the differences between the four distributions were statistically significant from each other at a level of less than $1\%$. This suggests that the human physical activity has unique characteristics for each of these situations.

\subsection{The Burstiness parameter}
The burstiness parameter $B_n$ in Eq.~(\ref{eq:bn}) was proposed as a value that characterises the shape of the IET distribution~\cite{Goh_2008, PhysRevE.94.032311}. It takes a value range of -1 to 1, with higher values indicating higher burstiness, that is, a larger standard deviation of the IETs compared to the mean. This indicator $B_n$  takes into account the effect of IETs smaller than $\tau_{\rm min}$, where the scaling behaviour begins, which is ignored in the discussion of the IETs distribution shape in the previous section. The burstiness parameters for each situation are shown in Table~\ref{table:result}. Among the four situations, \textit{Rest} has the highest burstiness, indicating that it is a behaviour with significant temporal fluctuations.

\begin{table}[ht]
\caption{Fitting results with the truncated power-law and the burstiness parameters for each situation. $\tau_{\rm min}$ is the point where the scaling behaviour begins, and $\alpha$ and $\lambda$ are the parameters of the truncated power-law.}
      \begin{tabular}{c|ccc|c}
        \hline
         & \multicolumn{3}{c|}{truncated power-law fit} & Burstiness \\ 
        situations   & $\tau_{\rm min}~[\si{\milli\second}]$ & $\alpha$ & $\lambda$ & $B_n$ \\  \hline
        Children's play & 620 & 1.68 & 1.86e-05 & 0.56 \\
        Housework & 574 & 1.60 & 1.41e-05 & 0.53  \\
        Desk work & 307 & 1.41 & 2.77e-06 & 0.57  \\ 
        Rest & 784 & 1.73 & 3.10e-06 & 0.68  \\ 
        \hline
      \end{tabular}
      \label{table:result}
\end{table}

\section{Discussion}
All the physical activities observed in this study, including that of children, shows a long-tailed IET distribution. This ﬁnding is essential for investigating the origin of burstiness because it suggests that burstiness in human behaviour is not acquired as humans mature. There are no existing studies on burstiness in children's activity. 

The IET distributions showed truncated power-law distributions in all situations. This means that the IETs have a scaling behaviour in certain value ranges, while the 90-min. observation period limits the tail.
There were statistically significant differences in IET distribution for each situation.
This unique curves for each situation are an important finding from the following two perspectives. First, this unique curve suggests that the IETs long-tailed distributions of human physical activity shown in previous studies~\cite{PhysRevLett.99.138103, PICOLI2022127160, PhysRevE.102.012307} are not caused by a Poisson process that changes the rates at which they perform an event over time~\cite{HIDALGOR2006877}. A human physical activity usually includes several different situations, and the Poisson process could produce IETs long-tailed distributions where each situation has a different constant event rate. However, the unique curves obtained in this study demonstrate that each situation has a different burstiness as unique characteristics. Second, this unique curve represents the temporal pattern of momentary thoughts and tentative breaks in the situation. A previous study on human behavioural inhibition~\cite{logan1984ability} demonstrated that the delay between the occurrence of a stop signal and the response takes approximately $300~\si{\milli\second}$. In this study, the time scale of the targeted IETs ranged from approximately $300~\si{\milli\second}$ to $100~\si{\second}$, and this scale may be related to momentary thoughts and tentative breaks rather than physiological reaction times.

According to previous studies, the power-law exponent $\alpha$ of IET distributions in individual activities ranges from approximately 0.7-2~\cite{karsai2018bursty}. The values observed in this study fall into this range. $B_n$ falls within the range of values reported for human activity data in previous studies \cite{Goh_2008}. 
Among the four situations, \textit{Desk work} has the smallest $\alpha$ and the longest tail distribution. This is a reasonable result, because during \textit{Desk work}, the physical activity level is expected to be low and IETs are long. On the other hand, $B_n$ is larger during \textit{Rest} than during \textit{Desk work}. In general, $B_n$ is larger due to the presence of very small IETs as well as the presence of very large IETs. The large $B_n$ during \textit{Rest} may be due to the frequency of IETs that are smaller than $\tau_{\rm min}$ rather than the tail of the IET distribution.


This study particularly focuses on human physical activity, which has two main advantages. First, the children's activities can be included in the study and analysed using the same methodology as adult activities. Children's activities are typically excluded from most human activity data used in previous studies (such as emailing and library loans).
Second, when measuring the physical activity, the individual's attributes and the situation can be recorded together, and the relationship between these and the statistical characteristics of the activity can be analysed. Using only web-based activity logs can be challenging to conduct such analysis while also identifying the subjects' situations.

Focusing on physical activity has the above advantages, but it also has challenges. First, the heavy load cannot be applied to a child in the observation. Therefore, in this study, a light accelerometer was attached to the torso, and measurements were taken continuously for 90 min. It has generally been noted that observation bias problems may occur at ﬁnite observation window for event time series~\cite{PhysRevE.92.052813}, and it is necessary to be concerned about the magnitude of the bias effect on the IET distribution at the time scale of interest, particularly for short observation windows.
In this study, we follow the guidelines proposed by~\cite{PhysRevE.92.052813} for bias estimation at the time scale of interest during the 90 min. observation window. 
Assuming that the observed event series was produced by the simplest model, the stationary renewal process, the observe IET distributions $P(\tau)$ of \textit{Children's play} and \textit{Housework} could be estimated to be at most $3.6\%$ smaller than the real distributions. The observed distributions $P(\tau)$ of \textit{Desk work} and \textit{Rest} could be estimated to be at most $16\%$ and $19\%$ smaller, respectively. This effect of observation bias is proportional to the size of $\tau$. For a 90-min. observation, for example, if we focus only on $\tau$ less than $54~\si{\second}$, the effect of bias is less than $1\%$.

The second difficulty is how to obtain an activity event series from an activity volume time series. 
In this study, a fixed threshold $A_c$ was given for activity amount, and activity amounts above the threshold were considered as activity events. Additionally, we confirm that the main conclusions are independent of the threshold, except for certain value ranges (see APPENDIX).

Finally, we were unable to discuss whether there are individual differences in the distribution of IETs. To achieve this, we must collect a larger sample and consider less time-consuming measurement methods.
Generally, there is a trade-off between offline and online observations in terms of data quality and quantity, and it is necessary to devise ways to obtain insights from both. 
For example, click behaviour on the web can be regarded as a series of behavioural events with a single purpose, and it is relatively easy to secure a large number of samples. On the other hand, offline observations like ours need more careful setups and efforts.
It will be interesting future research to investigate whether the unique curves for each type of behavior that characterise the thinking and resting time obtained in this study can be confirmed for click behaviour on the web. Furthermore, there is a need to investigate the theoretical model that generates these curves.


\begin{backmatter}

\section*{Availability of data and materials}
Data sufficient to reproduce all results in this paper will be made available upon request.

\section*{Funding}
Not applicable.

\section*{Abbreviations}
IET,Inter-Event Time.

\section*{Competing interests}
YS has no conficts of interest to declare. MT are employees of CyberAgent, Inc. There are no patents to declare.

\section*{Author's contributions}
MT and YS performed the study design, and the data collection. 
MT analyzed the data. MT and YS wrote the manuscript.
All authors read and approved the final manuscript.

\section*{Acknowledgements}
The authors thank Dr. Mai Otsuki and Dr. Yuki Hashimoto for their advice in selecting the sensor and other observational equipment. We also thank Mr. Chung Ming Hui for English advice.


\bibliographystyle{bmc-mathphys} 
\bibliography{bmc_article}      









\end{backmatter}

\appendix

\section{Confirmation of threshold dependence}
\label{sec:appendix_a}
This appendix provides a confirmation of the dependence of this study's conclusions on the activity threshold $A_c$. We examine the IET distribution when $A_c$ is varied for converting the activity time series $A(t)$ into an event sequence. In the study, $A_c$ is set at $100~\si{\metre\cubic\per\second}$, and then we demonstrate the effects of varying it from $30~\si{\metre\cubic\per\second}$ to $300~\si{\metre\cubic\per\second}$.

Generally, if the threshold is too low, the effect of random noise increases and is close to the exponential distribution of a stationary Poisson process. On the other hand, if the threshold is too high, the number of events becomes small and the shape of the distribution becomes unclear.

\begin{figure}[ht]
\centering
 \includegraphics[scale=0.4]{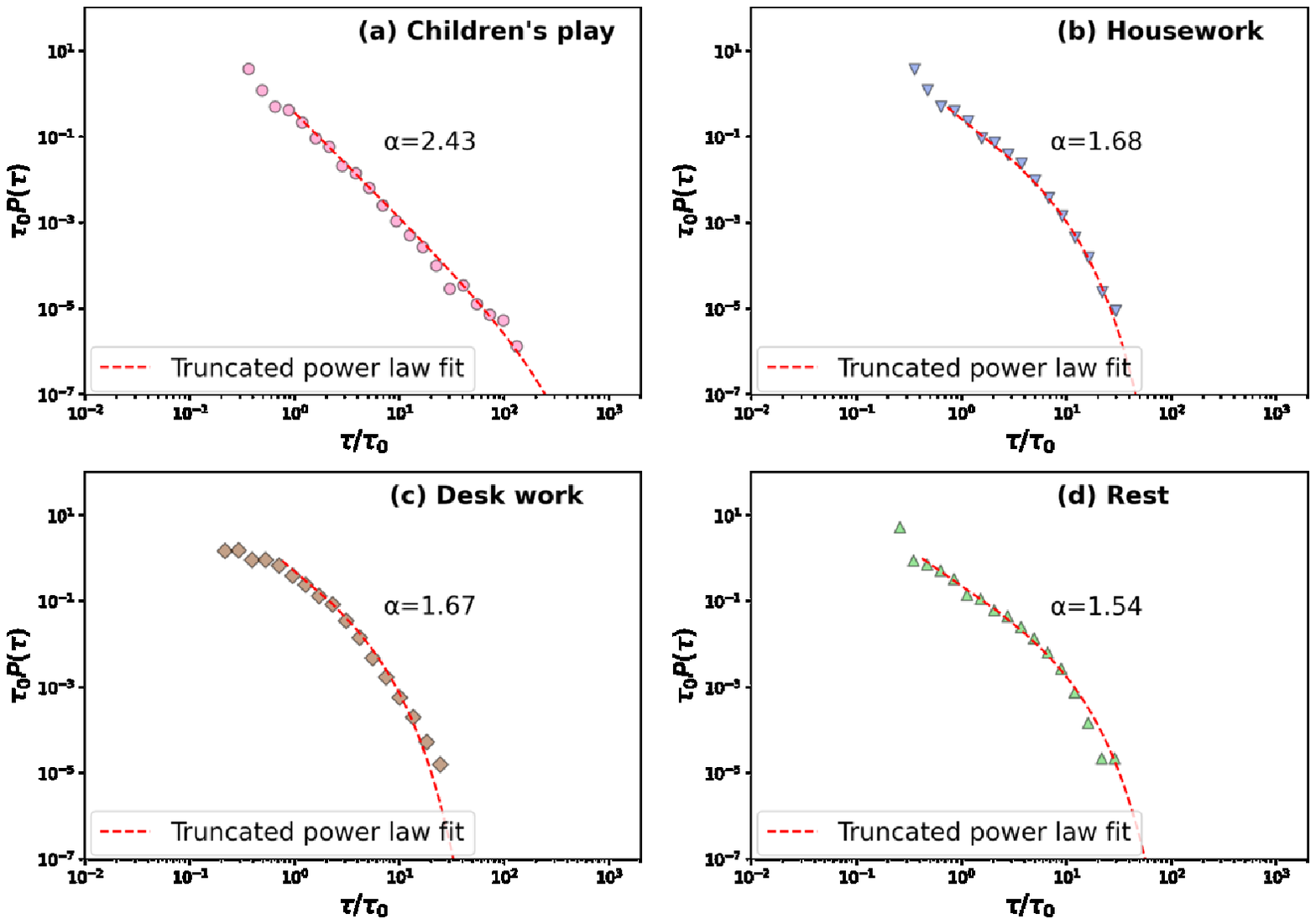}
 \caption{IET distributions with a threshold value of $30~\si{\metre\cubic\per\second}$.}
 \label{fig:iets_thre30}
\end{figure}

\begin{figure}[ht]
\centering
 \includegraphics[scale=0.4]{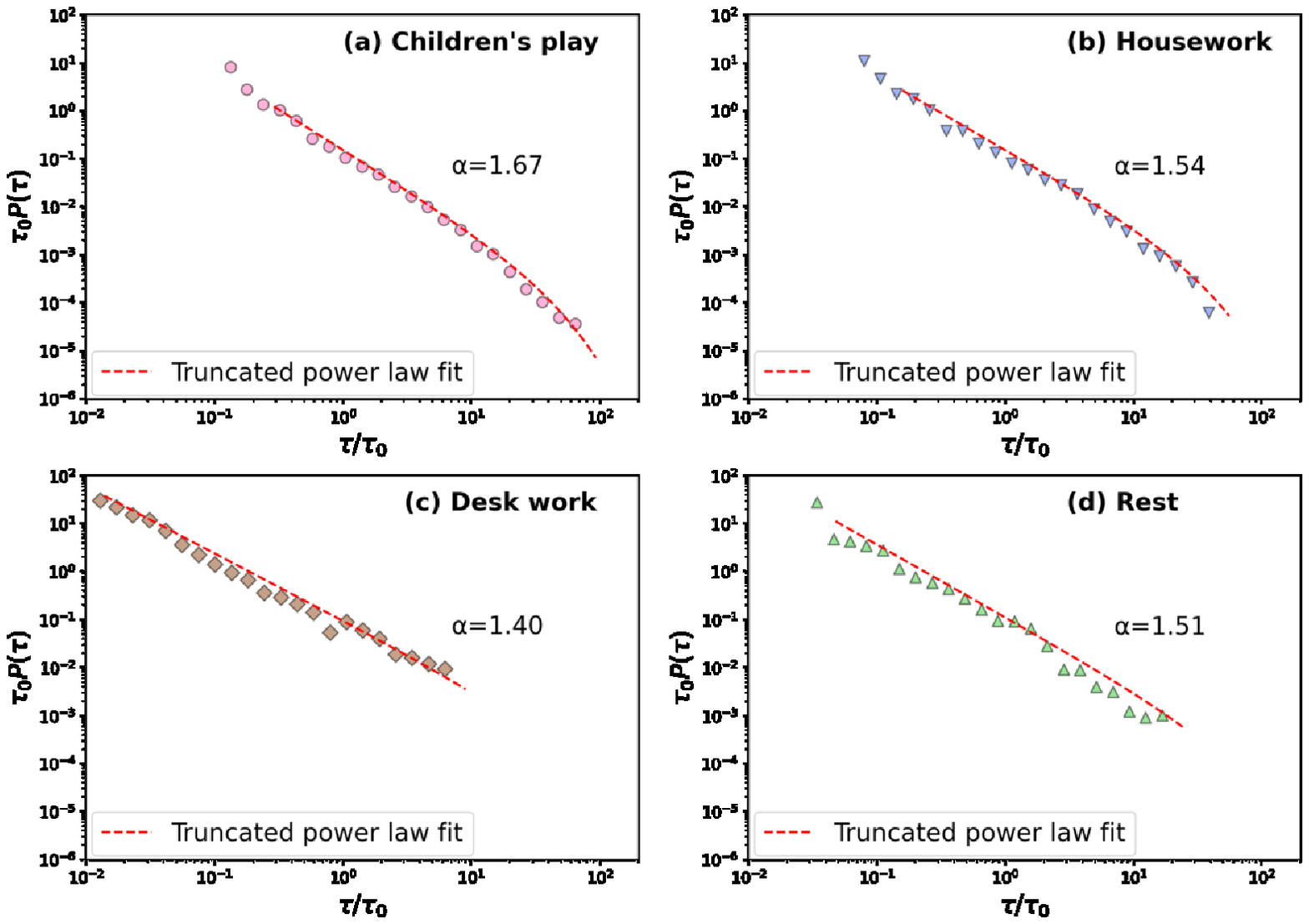}
 \caption{IET distributions with a threshold value of $150~\si{\metre\cubic\per\second}$.}
 \label{fig:iets_thre150}
\end{figure}

\begin{figure}[ht]
\centering
 \includegraphics[scale=0.4]{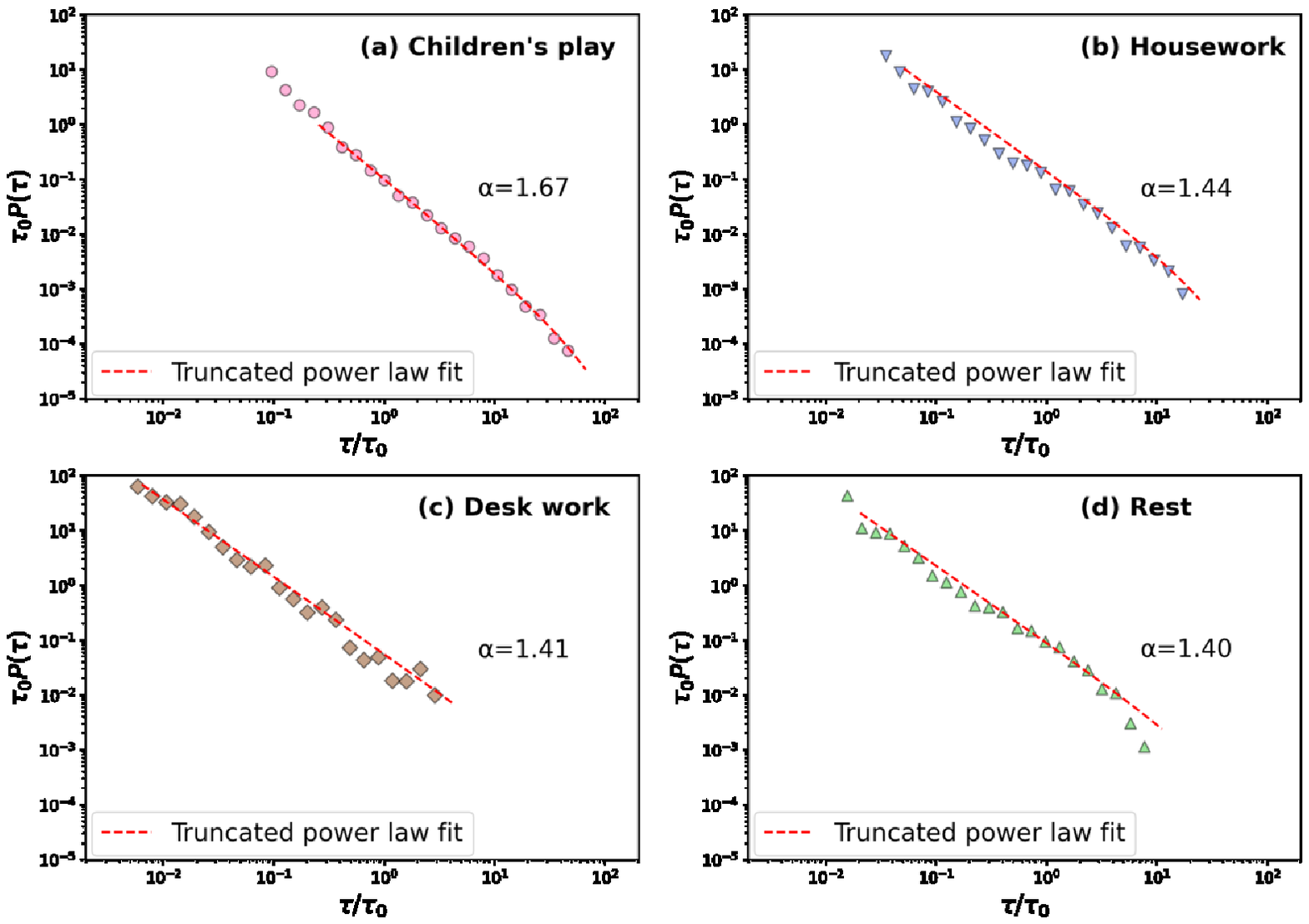}
 \caption{IET distributions with a threshold value of $300~\si{\metre\cubic\per\second}$.}
 \label{fig:iets_thre300}
\end{figure}
For $30~\si{\metre\cubic\per\second} \le A_c \le 100 ~\si{\metre\cubic\per\second}$, the long tail gradually shortens owing to random noise as the threshold value decreases. Figure~\ref{fig:iets_thre30} shows the distribution for the threshold value of 30~\si{\metre\cubic\per\second}.

For $100~\si{\metre\cubic\per\second} \le A_c \le 300 ~\si{\metre\cubic\per\second}$, all IET distributions had long tails.
As for the distribution function, except for \textit{Rest}, the conclusion remained the same: truncated power-law is the optimal function and the distributions for each situation had statistically significant differences from each other. For the distribution of \textit{Rest}, no distribution function was identified that demonstrated a statistically significant strong fit between thresholds $150 ~\si{\metre\cubic\per\second}$ and $300 ~\si{\metre\cubic\per\second}$, and truncated power-law and lognormal and stretched exponential are possible fits. Additionally, no statistically significant differences were identified between the \textit{Housework} and \textit{Rest} distributions between thresholds $250 ~\si{\metre\cubic\per\second}$ and $300 ~\si{\metre\cubic\per\second}$.

For the Burstiness $B_n$, the values vary slightly with threshold value, but the trend of the highest burstiness for \textit{Rest} among the four situations remains constant for threshold values above $100~\si{\metre\cubic\per\second}$ (Table.~\ref{table:bn}).
\begin{table}[h!]
\caption{Values of the Burstiness $B_n$ when the threshold is varied from $30~\si{\metre\cubic\per\second}$ to $300 ~\si{\metre\cubic\per\second}$.}
      \begin{tabular}{c|cccc}
        \hline
         & \multicolumn{4}{c}{Burstiness $B_n$}  \\ 
        situations   & $A_c =30~\si{\metre\cubic\per\second}$  & $A_c =100~\si{\metre\cubic\per\second}$ & $A_c =150 ~\si{\metre\cubic\per\second}$ & $A_c =300 ~\si{\metre\cubic\per\second}$ \\  \hline
        Children's play & 0.52 & 0.56 & 0.59 & 0.65 \\
        Housework & 0.14 & 0.53 & 0.60 & 0.61 \\
        Desk work & 0.07 & 0.57 & 0.57 & 0.59 \\ 
        Rest & 0.21 & 0.68 & 0.69 & 0.75 \\ 
        \hline
      \end{tabular}
      \label{table:bn}
\end{table}

\end{document}